\documentclass[%
 reprint,
nofootinbib,
 amsmath,amssymb,
 aps,
]{revtex4-2}

\usepackage{graphicx}% Include figure files
\usepackage{dcolumn}% Align table columns on decimal point
\usepackage{bm}% bold math
\usepackage{amsmath,amsfonts,amssymb}
\usepackage{subfig}
\usepackage{tikz}
\usepackage{float}
\usepackage{appendix}
\usetikzlibrary{quantikz}
\usetikzlibrary {angles}
\usepackage{xcolor,soul,todonotes}
\usepackage{easyReview}
\begin{document}

\preprint{APS/123-QED}

\title{Boosted Imaginary Time Evolution of Matrix Product States}% Force line breaks with \\

\author{Benjamin C. B. Symons}
\affiliation{%
 The Hartree Centre \\ STFC, Sci-Tech Daresbury \\ Warrington WA4 4AD \\ United Kingdom
}%
\author{Dilhan Manawadu}
\affiliation{%
 The Hartree Centre \\ STFC, Sci-Tech Daresbury \\ Warrington WA4 4AD \\ United Kingdom
}%
\author{David Galvin}
\affiliation{%
 The Hartree Centre \\ STFC, Sci-Tech Daresbury \\ Warrington WA4 4AD \\ United Kingdom
}%
\author{Stefano Mensa}
 \email{stefano.mensa@stfc.ac.uk}
\affiliation{%
 The Hartree Centre \\ STFC, Sci-Tech Daresbury \\ Warrington WA4 4AD \\ United Kingdom
}%

\date{\today}% It is always \today, today,

\begin{abstract}
In this work, we consider the imaginary time evolution of matrix product states. We present a novel quantum-inspired classical method that, when combined with time evolving block decimation (TEBD), is able to potentially speed-up the convergence to a ground state compared to TEBD alone. Our method, referred to as boosted imaginary time evolution, relies on the use of reflections to boost to lower energy states. Interleaving TEBD steps with boosts reduces the total number of TEBD steps and potentially the computational cost required to imaginary time evolve a matrix product state to a ground state. We give the mathematical details of the method followed by an algorithmic implementation and finally some results for a simple test case.
\end{abstract}

\maketitle

%\tableofcontents

\section{Introduction}

Understanding interacting quantum many-body systems is of key interest in physics. Simulating these systems even in seemingly simple cases can quickly become computationally intractable as the amount of information that must be stored and processed grows exponentially with system size. Tensor networks have emerged as a powerful tool to efficiently approximate quantum states in certain regimes \cite{Orus2019TNrev,Bridgeman2017Hand}. For example, Matrix product states (MPS) are able to efficiently and accurately approximate the state of 1D quantum systems that exhibit low entanglement \cite{Fannes1992FiniteCorr,Ostlund1995Thermo, Verstraete2006mps-gs,Perez2007MPSrep,Hastings2007Area}. A common task in the study of quantum many-body systems is to prepare and estimate the energy of a ground state. It is well known that finding the exact ground state of a k-local Hamiltonian for $k\geq2$ is NP-hard (in fact it is QMA-complete) \cite{Bookatz2013QMA}. Despite this, there are many approximate methods that aim to efficiently find approximate ground states e.g. density matrix renormalisation group (DMRG) \cite{DMF1992White,DMA1993White}. 

In this work we will focus on the preparation of ground states of 1D nearest-neighbour quantum systems, represented by matrix product states, using imaginary time evolution (ITE). In principle ITE can be applied in an exact setting, in which case the computational cost is exponential in both space and time.  However, an MPS can be approximately imaginary time evolved using time evolving block decimation (TEBD) \cite{MPS2003Vidal}. In this paper we will detail a novel `quantum-inspired' heuristic method that, when combined with TEBD, can reduce the computational cost of imaginary time evolving an MPS to the ground state compared to TEBD alone. We begin by giving some background on imaginary time evolution and matrix product states before describing our method. 

\vspace{15mm}
A quantum system with Hamiltonian $\hat{H}$ is evolved through imaginary time $\tau = it$ under the action of the non-unitary ITE operator, $\hat{T} = e^{-\hat{H}\tau}$. Note that imaginary time is an un-physical concept but is nonetheless a useful mathematical tool. Applying the ITE operator to a state gives,

\begin{equation}
    \ket{\psi(\tau)} = A(\tau)e^{-\hat{H}\tau} \ket{\psi(0)},
\label{ITE_deriv_1}
\end{equation}

where $A(\tau) = \bra{\psi(0)} e^{-2\hat{H}\tau}\ket{\psi(0)}^{-1/2}$ is a prefactor required to ensure normalisation. Expanding Equation \ref{ITE_deriv_1} in terms of eigenstates of $\hat{H}$ gives,

\begin{equation}
    \ket{\psi(\tau)}
    = A(\tau) \sum_{j}e^{-E_j\tau} a_j\ket{\phi_j},
\label{ITE_deriv_final}
\end{equation}

where $E_j$ is the energy of the eigenstate $\ket{\phi_j}$. The right-hand side of Equation \ref{ITE_deriv_final} provides the crucial insight as to why imaginary time evolution is useful. The coefficient of each state $\ket{\phi_j}$ decays exponentially in imaginary time. The rate of decay is determined by the energy of the state. The higher the energy of the state, the more quickly it decays. As such, the lowest energy state will decay the slowest. Assuming that the initial state $\ket{\psi(0)}$ has a non-zero overlap with the ground state then, in the limit $\tau \rightarrow \infty$, only the ground state will remain i.e. $\ket{\psi(\tau\rightarrow\infty)} = \ket{\phi_0}$ where $\ket{\phi_0}$ is the ground state.

Matrix product states are a useful way of approximately representing a quantum state efficiently. Consider a 1D quantum lattice system with $N$ sites, each with site dimension $d$. A general pure state $\ket{\psi}$ can be written as follows,

\begin{equation}
    \ket{\psi} = \sum_{i_1 i_2 ... i_n}^{d} C_{i_1 i_2 ... i_N} \ket{i_1} \otimes \ket{i_2} \otimes ... \otimes \ket{i_N},
\label{mps1}
\end{equation}

The tensor $C$ in Equation \ref{mps1} grows exponentially with the number of sites. However, it is possible to approximate $C$ as a product of lower rank tensors that can be obtained by Schmidt decomposition. A matrix product state is written, 

\begin{equation}
    \ket{\psi} = \sum_{i_1 i_2 ... i_N}^{d} \text{Tr} \left( A^{i_1} A^{i_2} ...\ A^{i_N} \right) \ket{i_1 i_2 ... i_N},
\end{equation}

where the matrices $A^{i_n}$ are of maximum dimension $\chi\times\chi$ and $\chi$ is a parameter known as the bond dimension. While an MPS can be an exact representation of a state, the true usefulness of matrix product states is as a tool for approximately representing quantum states. Provided a given state is only weakly entangled (specifically it obeys a 1D area law), an approximate MPS can be made arbitrarily close to the true state by increasing $\chi$, with only polynomial cost \cite{Schuh2007MPSentangle}.

An MPS can be (imaginary) time evolved using time evolving block decimation \cite{Paeckela2019TimeEv}. Given a 1D nearest neighbour Hamiltonian, the core idea of TEBD is to Trotterise the time evolution operator and exploit the locality of the Hamiltonian such that the MPS can be updated using only local operations. This approach avoids the full exponential cost that would otherwise be associated with time evolving a state. A 1D nearest neighbour Hamiltonian, can be decomposed into even and odd terms $\hat{H} = \hat{H}_{even} + \hat{H}_{odd}$ where the even (odd) terms act on even (odd) indexed sites $j$ and a neighbour $j+1$. While $\hat{H}_{even}$ and $\hat{H}_{odd}$ do not commute, each of the terms contained within the even (odd) sum mutually commute. The resulting first order Trotterised evolution operator is given by,

\begin{equation}
    U(\Delta\tau) = e^{-\hat{H}_{odd}\Delta\tau}e^{-\hat{H}_{even}\Delta\tau},
\label{tebd}
\end{equation}

where $\Delta\tau$ is a step in imaginary time. An evolution of duration $\tau = N_s\Delta\tau$ can be achieved by applying the operator $N_s$ times i.e., $U(\Delta\tau)^{N_s}$. Equation \ref{tebd} contains only local terms and can therefore be implemented efficiently.

\section{Boosted Imaginary Time Evolution}

In this paper we present a novel quantum-inspired classical algorithm that combines the concept of amplitude amplification \cite{Grover} with imaginary time evolution. We refer to this method as `boosted imaginary time evolution' (BITE) because it augments standard imaginary time evolution with a sequence of `boosts’ that jump from a given state to a lower energy state. The boosts are achieved by a sequence of reflections similar to those used in amplitude amplification. We will begin by illustrating the method in a simple case before giving a more general picture. We then develop an efficient method of implementing boosts that avoids the need to explicitly perform reflections.

We begin by considering a 2D Hilbert space $\mathcal{H}_2$ spanned by $\{ \ket{\phi_0}, \ket{\phi_1} \}$ with corresponding energies $E_0 < E_1$. For simplicity we initialise the system in a uniform superposition $\ket{s} = \frac{1}{\sqrt{2}}(\ket{\phi_0} + \ket{\phi_1})$ but, we note that the method generalises to an arbitrary initial state. Evolving the system forward in imaginary time by a step $\Delta\tau$ results in a state $\ket{t} = e^{-\hat{H}\Delta\tau} \ket{s}$. Expanding out $\ket{t}$ gives,

\begin{equation}
    \ket{t} = \frac{A(\Delta\tau)}{\sqrt{2}}\left( e^{-E_0\Delta\tau}\ket{\phi_0} + e^{-E_1\Delta\tau}\ket{\phi_1}\right).
\label{t-state expand}
\end{equation}

It is evident from Equation \ref{t-state expand} that $\ket{t}$ is closer to the ground state than $\ket{s}$ i.e. $|\braket{t}{\phi_0}|^2 > |\braket{s}{\phi_0}|^2$. After a single imaginary time step this difference may be small (assuming the step size is small). Despite this, the difference can now be boosted by performing a series of reflections, inspired by amplitude amplification. 

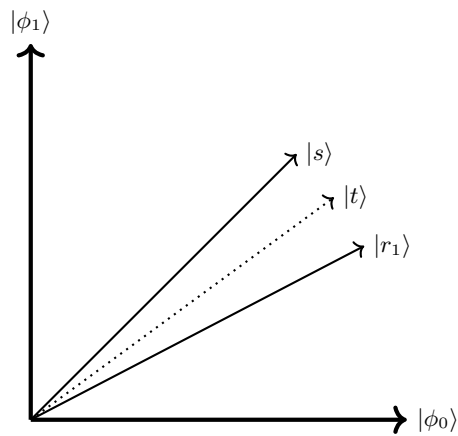
\begin{figure}[H]
\centering
\begin{tikzpicture}[scale=5.0]
    \draw[->, ultra thick](0,0) -- (1,0) node (xaxis) [right] {$\ket{\phi_0}$};
    \draw[->, ultra thick](0,0) -- (0,1) node (yaxis) [above] {$\ket{\phi_1}$};
    \draw[->, thick](0,0) -- (0.70712,0.70712) node (sp) [right] {$\ket{s}$};
    \draw[->, dotted, thick](0,0) -- (0.806225,0.591609) node (t) [right] {$\ket{t}$};
    \draw[->, thick](0,0) -- (0.8866676622700842,0.462407070057131) node (r) [right] {$\ket{r_1}$};
    \draw[] (0,0) coordinate node (origin) [right] {};
\end{tikzpicture}
\caption{The state $\ket{r_1}$ is  produced by reflecting $\ket{s}$ about $\ket{t}$.}
\label{const-1}
\end{figure}

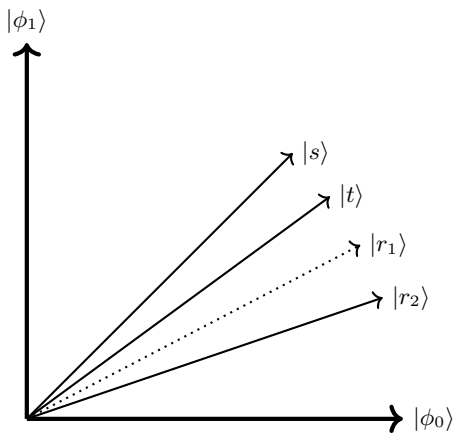
\begin{figure}[H]
\centering
\begin{tikzpicture}[scale=5.0]
    \draw[->, ultra thick](0,0) -- (1,0) node (xaxis) [right] {$\ket{\phi_0}$};
    \draw[->, ultra thick](0,0) -- (0,1) node (yaxis) [above] {$\ket{\phi_1}$};
    \draw[->, thick](0,0) -- (0.70712,0.70712) node (sp) [right] {$\ket{s}$};
    \draw[->, thick](0,0) -- (0.806225,0.591609) node (t) [right] {$\ket{t}$};
    \draw[->, dotted, thick](0,0) -- (0.8866676622700842,0.462407070057131) node (r) [right] 
    {$\ket{r_1}$};
    \draw[->, thick](0,0) -- (0.9465712361839997,0.3224937765757464) node (b) [right] {$\ket{r_2}$};
    \draw[] (0,0) coordinate node (origin) [right] {};
\end{tikzpicture}
\caption{The state $\ket{r_2}$ is  produced by reflecting $\ket{t}$ about $\ket{r_1}$.}
\label{const-2}
\end{figure}

The boosted imaginary time evolution procedure entails a `leapfrogging' of states using reflections as shown in Figures \ref{const-1} and \ref{const-2}. Firstly, an initial state $\ket{\psi_i}$ is imaginary time evolved by $N_s$ steps, $\ket{t} = U(\Delta\tau)^{N_s}\ket{\psi_i}$, where $U(\Delta\tau)$ is the evolution operator. A reflected state $\ket{r_1}$ is then produced by reflecting the initial state $\ket{\psi_i}$ about the evolved state $\ket{t}$ i.e., $\ket{r_1} = R(\ket{t})\ket{\psi_i}$. The unitary operator that implements a reflection about a state $\ket{v}$ is given by,

\begin{equation}
    R(\ket{v}) = 2\ket{v}\bra{v} - I.    
\label{reflection-operator}
\end{equation}

Figure \ref{const-1} shows the 2D case where the initial state is the uniform superposition state, $\ket{\psi_i} = \ket{s}$. Note that the larger $N_s$, the larger the angle of the reflection. In the second step, the state $\ket{t}$ is reflected about $\ket{r_1}$ to produce $\ket{r_2}$ (see Figure \ref{const-2}) and this can continue as many times as desired with the $n^{\text{th}}$ state given by the recurrence relation,

\begin{equation}
    \ket{r_n} = R(\ket{r_{n-1}}) \ket{r_{n-2}},
\label{rec-rel}
\end{equation}

with the initial conditions $\ket{r_0}=\ket{t}$ and $\ket{r_{-1}}=\ket{\psi_i}$.

Using the simple 2D example in Figures \ref{const-1} and \ref{const-2} we can further develop the method and demonstrate how it generalises beyond 2D. In this example, we see that the initial states that are `input' to the BITE procedure, $\ket{\psi_i}=\ket{s}$ and $\ket{t}$, define a 2D plane in Hilbert space, with the allowed states lying on a unit circle in the plane. The sequence of reflection operations that produces the sequence of states $\ket{r_n}$ traverses the unit circle in a series of rotations that have angle $\theta = \arccos(\braket{s}{t})$. In the 2D case, the unit circle corresponds to all allowed states in the Hilbert space and it is therefore the case that the ground state lies on the unit circle defined by the sequence of reflections. Only finitely many states are accessible by performing reflections, specifically the number of states is $n=2\pi/\theta$. In general the ground state will not be one of the accessible states. However, the minimum energy state accessible by reflections $\ket{r_{min}}$ will be close to the ground state. More concretely, the angle between $\ket{r_{min}}$ and $\ket{\phi_0}$ will be at most $\theta' = \arccos(\braket{s}{t})/2$.

\begin{figure}[H]
    \centering
    \includegraphics[width=0.375\textwidth]{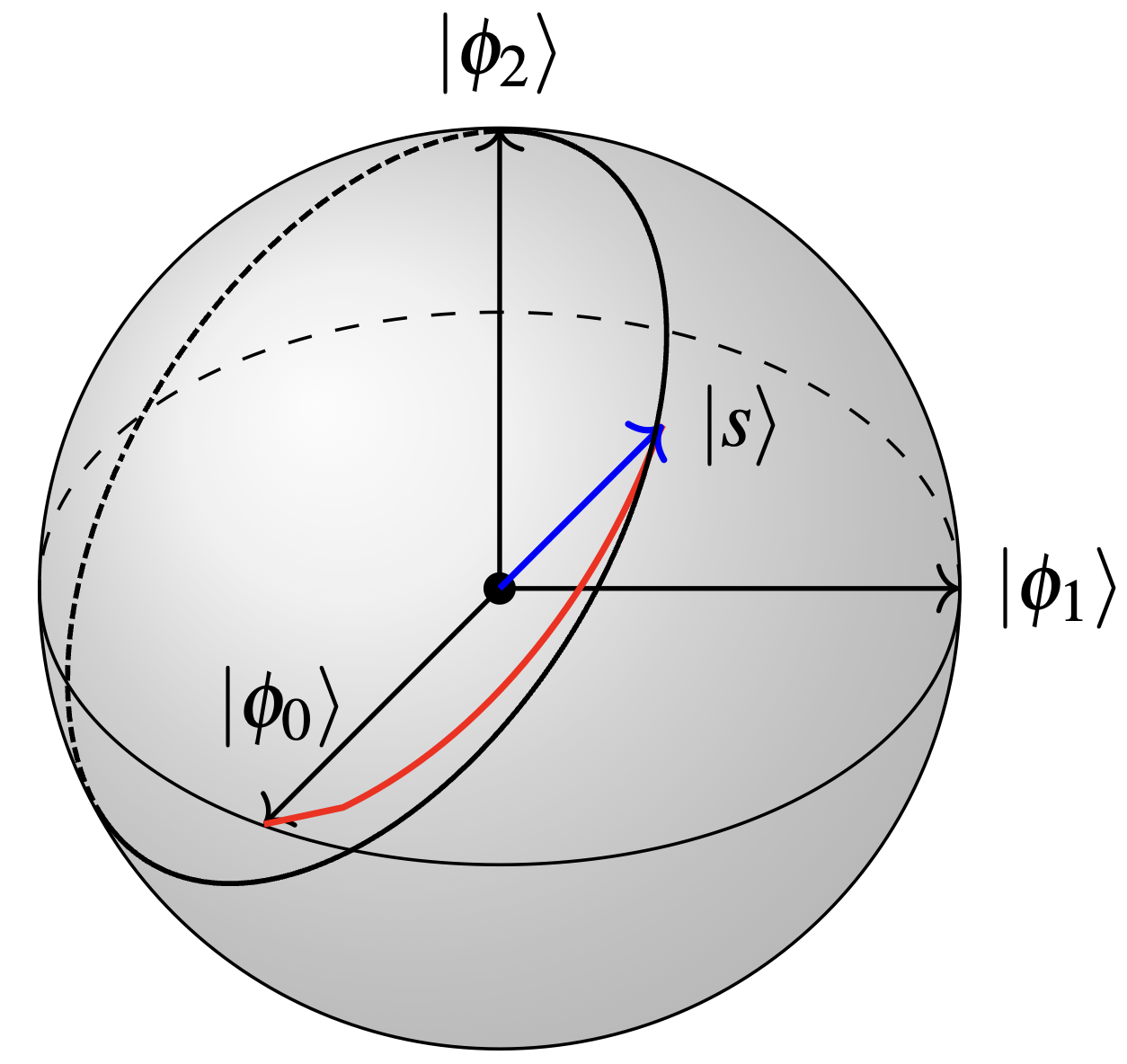}
    \caption{A visualisation of a 3D Hilbert space with the allowed states shown as the surface of a unit sphere. The blue vector represents the initial state $\ket{s}$ and the red curve is a trajectory resulting from imaginary time evolution. The black great circle that contains $\ket{s}$ is an example of the intersection of a plane with the surface of the unit sphere.}
    \label{fig:sphere}
\end{figure}

The 2D case is a somewhat trivial case and so we will now give an illustrative 3D example. Consider a 3D Hilbert space $\mathcal{H}_3$ spanned by $\{ \ket{\phi_0}, \ket{\phi_1}, \ket{\phi_2} \}$ with respective energies $E_0 < E_1 < E_2$. The allowed states $\ket{\psi} \in \mathcal{H}_3$ lie on the 2D surface of the unit 2-sphere. Without loss of generality consider the states $\ket{\phi_0} = (1,0,0)^T$, $\ket{\phi_1} = (0,1,0)^T$ and $\ket{\phi_2} = (0,0,1)^T$. Given some starting state $\ket{\psi_i}=(a_0,a_1,a_2)^T$, imaginary time evolution will trace out a path on the surface of the unit sphere defined by $A(\tau)(a_0 e^{-E_0\tau}, a_1 e^{-E_1\tau}, a_2 e^{-E_2\tau})^T$. The path will begin at $\ket{\psi_i}$ and, in the limit of large $\tau = N_s\Delta\tau$, will terminate at the ground state. An example trajectory with $\ket{\psi_i} = \ket{s}$ is shown by a red curve in Figure \ref{fig:sphere}.

Now consider the 2D plane defined by the states $\ket{s}$ and an imaginary time evolved state $\ket{t}$ (not shown in Figure \ref{fig:sphere}) which will intersect the unit sphere defining a great circle. The great circle defines a set of states $C$ such that $\ket{r_n} \in C$ are accessible by performing reflections. An example of such a great circle is shown by the black circle in Figure \ref{fig:sphere} that contains $\ket{s}$ and initially intersects the red ITE path. Performing a series of reflections will traverse this great circle, eventually returning to the initial state. It is evident that, unlike the 2D case, the ground state will not generally lie on the great circle, $\ket{\phi_0} \not\in C$. However, it can be qualitatively seen from Figure \ref{fig:sphere} that traversing $C$ until a minimum energy state is found $\ket{r_{min}}$ will result in a state that is closer to the ground state i.e. $|\braket{r_{min}}{\phi_0}|^2 > |\braket{t}{\phi_0}|^2$. To put this more rigorously, provided all states $\ket{r_n} \in C$ respect any physical constraints of the system, the minimum energy state $\ket{r_{min}}$ is guaranteed to be closer to the ground state provided $E_{r_{min}} < E_t$. We now develop an efficient way of performing boosts without needing to explicitly perform reflections and then go on to detail a potential algorithmic implementation of BITE.

\subsection{Recurrence Relations for Efficient Boosting}

One of the key features of the BITE method is that it can be implemented in a highly efficient manner by exploiting the recurrence relation in Equation \ref{rec-rel}. It can be shown (see Appendix \ref{App-1}) that the recurrence relation for the $n^{\text{th}}$ reflected state can be written as follows,

\begin{equation}
    \ket{r_n} = 2F\ket{r_{n-1}} - \ket{r_{n-2}},
\label{rr1}
\end{equation}

where we have defined the overlap $F=\braket{t}{\psi_i}$. It is then possible to use Equation \ref{rr1} to express the $n^{\text{th}}$ state in terms of only the input states and some coefficients $\alpha_n$ and $\beta_n$,

\begin{equation}
    \ket{r_n} = \alpha_n\ket{t} + \beta_n\ket{\psi_i}.
\label{rr2}
\end{equation}

Furthermore, we show in Appendix \ref{App-1} that the coefficients $\alpha_n$ and $\beta_n$ can be calculated using the following recurrence relations,

\begin{equation}
\begin{split}
    &\alpha_n = 2F\alpha_{n-1} - \alpha_{n-2}, \\
    &\beta_n = -\alpha_{n-1}, 
\end{split}
\label{rr-coeff}
\end{equation}

with the initial conditions $\alpha_{0} = 1$, $\alpha_{-1} = 0$ and $\alpha_{-2}=-1$. In order to reach the state $\ket{r_n}$ we no longer need to perform the full sequence of reflections. Instead we need only compute the series of coefficients $\alpha_n$ using the recurrence relations in Equation \ref{rr-coeff} and then apply Equation \ref{rr2}. This allows us to boost directly to the $n^{\text{th}}$ state without computing the intermediate states. In Appendix \ref{App-2} we derive a closed form expression for the $n^{\text{th}}$ state,

\begin{equation}
    \ket{r_n} = \frac{1}{\sin(\theta)}\left( 
    \sin((n+1)\theta) \ket{t} - \sin(n\theta)\ket{\psi_i}
    \right), 
\label{closed-form}
\end{equation}

where $\theta = \arccos(F)$. The closed form in Equation \ref{closed-form} enables an even more efficient boost. However, it is not necessarily apparent which state to boost to. In order to select a state to boost to, it would be useful to know the energy of the state ahead of time. Consider the energy of the $n^{\text{th}}$ state, $E_n = \bra{r_n} \hat{H} \ket{r_n}$. Using Equation \ref{rr2} we can derive a recurrence relation for the energy,

\begin{equation}
\begin{split}
    E_n = \alpha_n^2 E_t + \alpha_{n-1}^2E_i - 2\alpha_n\alpha_{n-1}E_{it}, 
\end{split}
\label{rr-En}
\end{equation}

where $E_t = \bra{t}\hat{H}\ket{t}$, $E_i = \bra{\psi_i}\hat{H}\ket{\psi_i}$ and $E_{it} = \bra{\psi_i}\hat{H}\ket{t}$. The recurrence relation in Equation \ref{rr-En} allows the energy of the $n^{\text{th}}$ state to be computed without computing the state itself. This means that the series $E_n$ can be evaluated efficiently in order to choose a state to boost to and then Equation \ref{closed-form} can be used to boost directly to the chosen state. We now have everything we need in order to describe our heuristic algorithm that implements the BITE method.

\subsection{Algorithm Implementation}

In this section we will describe a heuristic algorithm that implements boosted imaginary time evolution and in the following section, we demonstrate the algorithm on a simple test case of a 1D spin chain. The BITE algorithm sequentially performs many boosts, each time using new states to define a new 2D plane and therefore a new great circle on the surface of the n-sphere. The steps of the algorithm are as follows:
\vspace{5mm}
\begin{enumerate}
    \item Imaginary time evolve an initial state $\ket{\psi_i}$ by a single step $\Delta\tau$ to produce a state $\ket{t}$.
    \item Calculate the overlap $F$ and the terms $E_t$, $E_i$ and $E_{it}$ and use Equation \ref{rr-En} to compute the series of energies $E_n$ of states that lie on the great circle defined by the boost. Find the minimum $E_n$ and compute the corresponding state $\ket{r_n}$.
    \item Imaginary time evolve $\ket{r_n}$ by a single step to produce a new initial state $\ket{\psi_i'}$, and by two steps to produce a new evolved state $\ket{t'}$.
    \item Repeat steps 1-3 using the new states $\ket{\psi_i'}$ and $\ket{t'}$ as the initial states until the energy converges.
\end{enumerate}

The key advance that makes this method computationally viable is that, by using expression for the series of energies $E_n$ in Equation \ref{rr-En}, it is possible to find the minimum energy state on the plane efficiently. The computational cost associated with boosting from $\ket{t}$ to $\ket{r_n}$ is comprised of two parts: the cost to find the minimum $E_n$ and the cost to construct the associated state $\ket{r_n}$. Computing the series $E_n$ requires the overlap $F$ as well as the quantities $E_t$, $E_i$ and $E_{it}$. Computing each of these quantities can be done with an overlap which, to leading order, requires $O(2Nd\chi^3)$ operations for an MPS with $N$ sites and site dimension $d$ \cite{DMRGage2011Schollwock}. This gives a total of $O(8Nd\chi^3)$ operations to calculate the quantities needed to compute the series $E_n$. Computing the state $\ket{r_n}$ requires the coefficients $\alpha_n$ and $\alpha_{n-1}$ which can be computed with $O(1)$ cost given $F$ and then a single MPS add of the form $\ket{c} = \alpha\ket{a} + \beta\ket{b}$. An MPS addition operation is in principle no cost but, the direct sum grows the bond dimension from $\chi$ to $2\chi$. The computational cost is therefore associated with compressing the MPS after the addition. Assuming that we wish to compress that state back to a bond dimension of $\chi$ by SVD compression, the required operations are $O(4Nd\chi^3)$ \cite{DMRGage2011Schollwock}. The total cost for a boost is therefore $C_B \in O(Nd\chi^3)$. Note, there is also some space overhead because boosting requires storing 2 MPS ($\ket{\psi_i}$ and $\ket{t}$) rather than updating a single MPS in place.

The cost of implementing a single TEBD step is dominated by the cost of applying 2-body terms which is $C_T \in O(Nd^3\chi^3)$. We therefore see that, to leading order, the ratio of the cost of boosting and the cost of a single TEBD step is,

\begin{equation}
    \frac{C_B}{C_T} = \frac{K_B Nd\chi^3}{K_T Nd^3\chi^3} = \frac{K_B}{K_T d^2},
\label{cost-ratio}
\end{equation}

where $K_B$ and $K_T$ are the constant prefactors for boosting and TEBD respectively. Importantly, the cost ratio in Equation \ref{cost-ratio} is independent of system size and bond dimension. The constant factor $K_T$ contains a factor $f(p)$ that depends on the order of Trotterisation $p$ used. Higher order Trotterisation requires more operators to implement, thereby increasing the cost of a TEBD step. Given that we will consider spin systems, we fix the site dimension as $d=2$. The cost ratio is therefore a constant factor $K=K_B/4K_T \in O(1)$. A boost is therefore efficient provided it progresses further towards the ground state than $K$ TEBD steps would have i.e. $|\bra{\phi_0}U_{boost}\ket{\psi}|^2 > |\bra{\phi_0} U_{tebd}^K\ket{\psi}|^2$. It is evident that it is not \textit{a priori} possible to guarantee that the above inequality is satisfied and that boosting is efficient. In light of this fact, we found it useful to  introduce a heuristic that only begins boosting once the rate of change of the energy of the state evolved by TEBD falls below a certain threshold. Such a heuristic helps to reduce the likelihood that boosting is used in a regime where it is inefficient. Note that, as the order of Trotterisation increases, the ratio $K$ decreases and it is also evident that for smaller time step sizes $\Delta\tau$ the action of $U_{tebd}^K$ on a state $\ket{\psi}$ will move the state a smaller amount towards the ground state. We therefore expect boosting to more useful in cases with $p\geq 2$ and smaller $\Delta\tau$ i.e., where a smaller Trotter error is necessary.

\section{Results}

In this section we demonstrate the BITE algorithm for a simple test case of the transverse field Ising model on a 1D spin chain of $N=100$ sites, with open boundary conditions and parameters $J=0.5$ and $g=1.5$.\footnote{Other non-critical parameter values tested gave similar results to those below.} The initial state used is a product state of random superpositions of up and down spins. All calculations were run in serial on an Intel Xeon Gold 6142 CPU, using the python package TeNPy \cite{tenpy} which defines the transverse field Ising model as,

\begin{equation}
    \hat{H} = -J\sum_{i,j} X_i X_j -g \sum_{i}Z_i.
\end{equation}

\begin{figure}[H]
    \centering
    \includegraphics[width=0.49\textwidth]{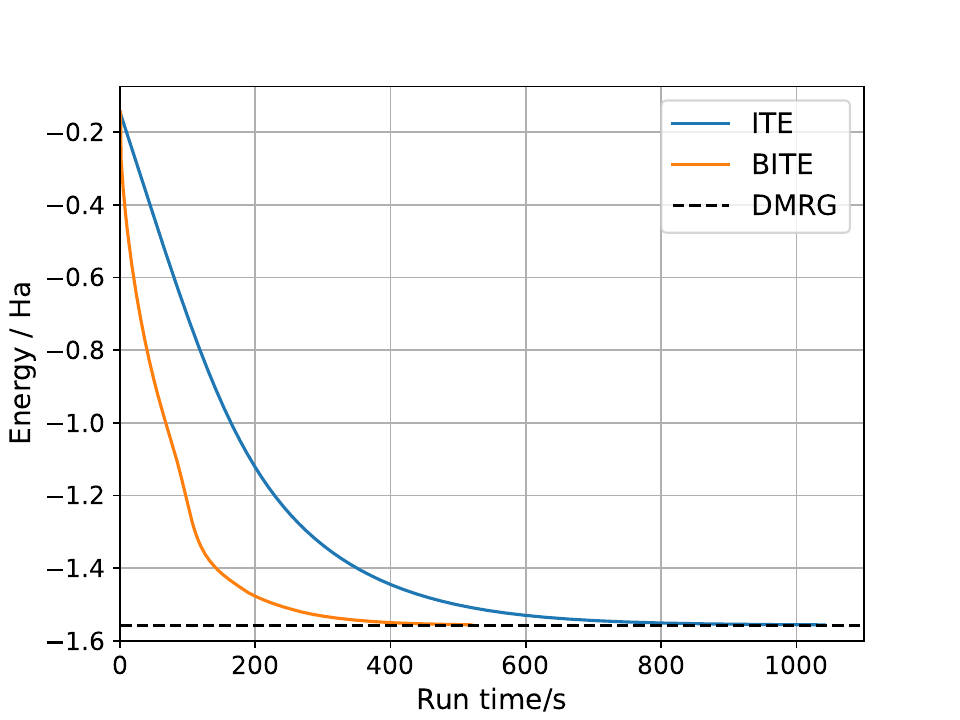}
    \caption{The energy per site of the MPS vs run time for ITE with 2$^{\text{nd}}$ order TEBD and the BITE algorithm. Calculations terminated when the energy was within 1mHa of the reference DMRG energy (dashed black line).}
    \label{fig:energy}
\end{figure}

\begin{figure}[H]
    \centering
    \includegraphics[width=0.49\textwidth]{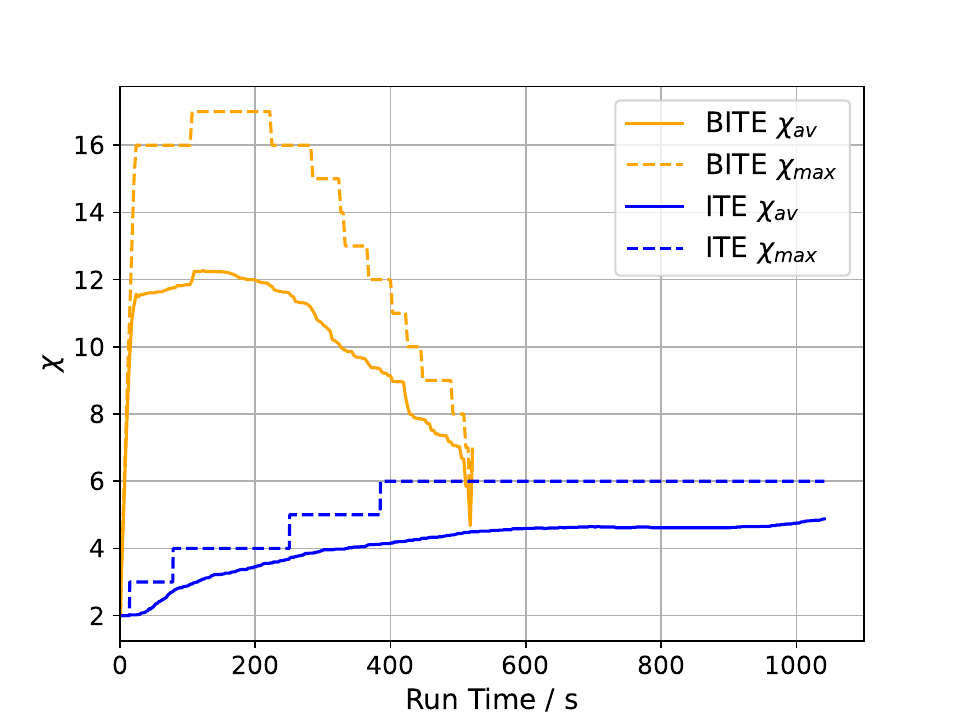}
    \caption{Average and maximum bond dimension at each point in the BITE and ITE calculations. $\chi_{av}$ is averaged over all matrices in the MPS at a given step.}
    \label{fig:chi}
\end{figure}

In Figure \ref{fig:energy} we compare standard imaginary time evolution using 2$^{\text{nd}}$ order TEBD with step size $\Delta\tau = 0.001$ to the boosted imaginary time evolution algorithm in the previous section. In Figure \ref{fig:energy} BITE converges to within 0.001 Ha of the density matrix renormalisation group energy in a run time of $t_r=521$s compared to $t_r=1042$s for ITE. The run time $t_r$ is the total wall time for the calculation. This is a considerable reduction in run time, especially considering the fact that we allow the bond dimension to grow somewhat during the MPS addition operations. In Figure \ref{fig:chi} we show $\chi_{av}$ and $\chi_{max}$ which are respectively, the mean and max bond dimensions at each point in the calculation. We see that, despite BITE having a $\chi_{max}$ of up to 17 compared to $\chi_{max}=6$ for ITE, the run time reduction is still significant. However, it is important to note that we used a small time step and this is favourable for the BITE algorithm as each TEBD step progresses a smaller distance towards the ground state.

In Table \ref{tab:table-1} we show how the performance changes as the time step increases from $\Delta\tau =0.001$ to $0.01$. With a larger time step, we find that it is not necessarily beneficial to begin boosting immediately as we did in Figure \ref{fig:energy}. This is because, with a larger time step, the initial part of the ITE trajectory improves the state rapidly and it is not necessarily the case that boosting improves the state faster than TEBD. We therefore implement a simple heuristic that determines when to start boosting; we start boosting once the gradient of the energy falls below a certain threshold, $|m_E| \leq T$ where $m_E = \Delta E/\Delta\tau$. In all cases shown in Table \ref{tab:table-1} we use $T=1$ (including for $\Delta\tau=0.001$ although in this case we find $|m_E| < 1$ from the start) and find reductions in run time of $\sim 25\%$ for both $\Delta\tau=0.005$ and $\Delta\tau=0.01$. We also show the fidelities of the final state with the DMRG ground state in Table \ref{tab:table-1} and find that BITE and ITE are comparable in most cases. However, in the case of $\Delta\tau=0.001$, the BITE fidelity is improved to 0.998 as compared to 0.953 for ITE. It is likely that in this case, the ground state was very close to being contained in the plane defined by the final boost thus leading to a considerable improvement in fidelity.

It is important to note that, for the simple system tested here, imaginary time evolution would not be the method of choice for finding the ground state. In this regime DMRG is significantly faster and is highly accurate. We present these results simply as a proof of principle to demonstrate that the BITE method can improve over imaginary time evolution using TEBD. Imaginary time evolution is used to study some systems \cite{Phien2015Recycle} and we anticipate that in some of those scenarios, our method may help to speed-up convergence to the ground state.

\begin{table}[h]
\caption{Run time, maximum and average (over the entire calculation) bond dimension and fidelity of the final state with the ground state for various time step sizes.}\label{tab:table-1}
\centering
    \begin{tabular}{|c|r|r|r|r|r|r|}
    \hline
    Method & $\Delta\tau$ & $t_r$ & $\chi_{max}$ & $\chi_{av}$ & $|\braket{\phi_0}{\psi}|^2$ \\
    \hline
    ITE & 0.001 & 1042 & 6  & 4.1 & 0.953 \\
    \hline
    BITE & 0.001 & 521 & 17 & 10.2 & 0.998\\
    \hline
    ITE & 0.005 & 204 & 7 & 4.5 & 0.956 \\
    \hline
    BITE & 0.005 & 156 & 16 & 6.3 & 0.959 \\
    \hline
    ITE & 0.01 & 105 & 7  & 4.8 & 0.960 \\
    \hline
    BITE & 0.01 & 79 & 16 & 5.8 & 0.957 \\
    \hline
    \end{tabular}
\end{table}

\section{Conclusion}

We have presented a novel quantum-inspired classical algorithm that, when combined with TEBD, has the potential to speed-up the convergence of imaginary time evolution of matrix product states to a ground state. We have shown how to use an initial and an imaginary time-evolved matrix product state to define a plane that intersects the unit n-sphere, defining a great circle that can be traversed by performing a sequence of reflections. By defining a state that has undergone a given number of reflections recursively, we were able to derive a closed form expression for any state (and its energy) on the great circle accessible by performing rotations. This closed form expression allows a so-called boosting procedure to be implemented efficiently.

We have presented a proof-of-concept demonstration of the method applied to the 1D transverse field Ising model with open boundary conditions. The BITE method was able to reduce the total run time to converge to within 1mHa of the DMRG energy by 25-50\% depending on the time step size. We find that BITE offers a larger improvement in regimes when a TEBD step only improves the state by a relatively small amount i.e. during the later part of an imaginary time evolution and when the time step is relatively small. The method may therefore be best suited to cases when a small Trotter error is required. In this work we have focused on the 1D case where DMRG is typically a better method than ITE. However, it may be possible to extend the BITE method to 2D where the use of imaginary time evolution is more prevalent. For example, projected entangled pair states (PEPS) \cite{Verstraete2004Peps} are typically updated using ITE \cite{Lubasch2014AlgorithmsPeps,Phien2015Recycle} and therefore might benefit from the BITE method.

\begin{acknowledgments}
This work was supported by the Hartree National Centre for Digital Innovation, a UK Government-funded collaboration between STFC and IBM.
\end{acknowledgments}

%\appendix

%\section{Appendixes}

%\nocite{*}

\bibliography{refs}% Produces the bibliography via BibTeX.

\begin{appendices}

\section{Derivation of Recurrence relations}\label{App-1}

The series of reflected states described in the main paper is defined by the recurrence relation $\ket{r_n} = R(\ket{r_{n-1}})\ket{r_{n-2}}$. Expanding the reflection operator of Equation \ref{reflection-operator} gives,

\begin{equation}
\begin{split}
    \ket{r_n} 
    &=
    R(\ket{r_{n-1}}) \ket{r_{n-2}}\\
    &=
    2\braket{r_{n-1}}{r_{n-2}}\ket{r_{n-1}} - \ket{r_{n-2}}.
\end{split}
\label{app-rr1}
\end{equation}

To compute the $n^{\text{th}}$ state, we require the overlap $\braket{r_{n-1}}{r_{n-2}}$. We will demonstrate using the recursive nature of the equations that this simply reduces to the overlap $\braket{t}{\psi_i}$,

\begin{equation}
\begin{split}
    \braket{r_{n-1}}{r_{n-2}} 
    &=
    \left( 2\braket{r_{n-3}}{r_{n-2}}\bra{r_{n-2}} - \bra{r_{n-3}} \right) \ket{r_{n-2}}\\
    &=
    2\braket{r_{n-3}}{r_{n-2}}- \braket{r_{n-3}}{r_{n-2}} \\
    &=
    \braket{r_{n-3}}{r_{n-2}}
\end{split}
\label{app-rr2}
\end{equation}

It is trivial to continue this process,

\begin{equation}
\begin{split}
    \braket{r_{n-3}}{r_{n-2}} 
    &=
    \bra{r_{n-3}} \left( 2\braket{r_{n-3}}{r_{n-4}}\ket{r_{n-3}} - \ket{r_{n-4}} \right)\\
    &=
    2\braket{r_{n-3}}{r_{n-4}}- \braket{r_{n-3}}{r_{n-4}} \\
    &=
    \braket{r_{n-3}}{r_{n-4}}
\end{split}
\label{app-rr3}
\end{equation}

Finally, consider repeating the process until reaching $\braket{r_2}{r_1}$,

\begin{equation}
\begin{split}
    \braket{r_{2}}{r_{1}} 
    &=
    \left( 2\braket{t}{r_1}\bra{r_1} - \bra{t} \right) \ket{r_1} \\
    &=
    \braket{t}{r_1} \\
    &=
    \bra{t} \left( 2\braket{t}{\psi_i}\ket{t} - \ket{\psi_i}\right) \\
    &= 
    \braket{t}{\psi_i}.
\end{split}
\label{app-rr4}
\end{equation}

We have therefore shown that $\braket{r_{n}}{r_{n-1}} = \braket{t}{\psi_i}$. This result should not be surprising because the reflection at each step is effectively a constant angle rotation. Using this result, we can rewrite the recurrence relation in Equation \ref{app-rr1} as,

\begin{equation}
    \ket{r_n} = 2\braket{t}{\psi_i}\ket{r_{n-1}} - \ket{r_{n-2}}.
\label{app-rr5}
\end{equation}

We now define $F = \braket{t}{\psi_i}$. Given that the state $\ket{r_n}$ is defined recursively, we can rewrite Equation \ref{app-rr5} in terms of the input states $\ket{\psi_i}$ and $\ket{t}$,

\begin{equation}
    \ket{r_n} = \alpha_n \ket{t} + \beta_n \ket{\psi_i},
\label{app-rr6}
\end{equation}

where $\alpha_n$ and $\beta_n$ obey their own recurrence relations. We will suppose the following form of Equation \ref{app-rr6},

\begin{equation}
    \ket{r_n} = \alpha_n \ket{t} - \alpha_{n-1}\ket{\psi_i},
\label{proof-1}
\end{equation}

with the recurrence relation,

\begin{equation}
    \alpha_n = 2F\alpha_{n-1} - \alpha_{n-2}
\label{proof-2}
\end{equation}

and initial conditions of $\alpha_0 = 1$, $\alpha_{-1} =0$, and $\alpha_{-2}=-1$. We will first demonstrate that Equations \ref{proof-1} and \ref{proof-2} hold for $n=1,2$ and then prove that they hold for all $n$ using a proof by induction. The states $\ket{r_1}$ and $\ket{r_2}$ are derived as follows using the recurrence relation in Equation \ref{app-rr5},

\begin{equation}
    \ket{r_1} = 2F\ket{t} - \ket{\psi_i},
\label{proof-3}
\end{equation}

\begin{equation}
\begin{split}
    \ket{r_2} &= 2F\ket{r_1} - \ket{t} \\
    &= 2F(2F\ket{t} - \ket{\psi_i}) - \ket{t} \\
    &= (4F^2 - 1)\ket{t} - 2F\ket{\psi_i}
\end{split}
\label{proof-4}
\end{equation}

We now derive the coefficients using the proposed recurrence relation in Equation \ref{proof-2},

\begin{equation}
    \alpha_1 = 2F\alpha_0 - \alpha_{-1} = 2F
\label{proof-5}
\end{equation}

\begin{equation}
\begin{split}
    \alpha_2 &= 2F\alpha_1 - \alpha_{0}\\
    &= 2F(2F) - 1\\
    &= 4F^2 - 1
\end{split}
\label{proof-6}
\end{equation}

It is evident that the proposed recurrence relation holds for $n=1,2$. Now suppose that the relation holds for $n=k$ and $n=k-1$,

\begin{equation}
    \ket{r_k} = \alpha_k \ket{t} - \alpha_{k-1} \ket{\psi_i}.
\label{proof-7}
\end{equation}

Now consider $n=k+1$ which according to Equation \ref{app-rr5} is given by,

\begin{equation}
    \ket{r_{k+1}} = 2F \ket{r_{k}} - \ket{r_{k-1}}.
\label{proof-8}
\end{equation}

Substituting the expressions for $\ket{r_k}$ and $\ket{r_{k-1}}$ gives,

\begin{equation}
\begin{split}
    \ket{r_{k+1}}
    =
    2F&(\alpha_k \ket{t} - \alpha_{k-1} \ket{\psi_i})\\
    & - (\alpha_{k-1} \ket{t} - \alpha_{k-2} \ket{\psi_i}),
\end{split}
\label{proof-9}
\end{equation}

\begin{equation}
\begin{split}
    \ket{r_{k+1}}
    =
    (2F\alpha_k - \alpha_{k-1})\ket{t} - (2F\alpha_{k-1} - \alpha_{k-2})\ket{\psi_i}.
\end{split}
\label{proof-10}
\end{equation}

Utilising the proposed recurrence relation in Equation \ref{proof-2} we see that the coefficient of $\ket{t}$ in Equation \ref{proof-10} is simply $\alpha_{k+1}$ and the coefficient of $\ket{\psi_i}$ is $\alpha_{k}$ and the expression therefore reduces to the expected form, thereby concluding the proof by induction,

\begin{equation}
    \ket{r_{k+1}} = \alpha_{k+1} \ket{t} - \alpha_{k} \ket{\psi_i}.
\label{proof-11}
\end{equation}

\section{Derivation of Closed Form}\label{App-2}

In this section we derive a closed form for the terms in the recurrence relation,

\begin{equation}
    \alpha_n = 2F\alpha_{n-1} - \alpha_{n-2},
\label{cf1}
\end{equation}

with initial conditions $\alpha_{-1} = 0$ and  $\alpha_0 = 1$. This is a homogeneous linear recurrence and we can therefore solve it using its characteristic equation,

\begin{equation}
    x^2 - 2Fx + 1 = 0.
\label{cf2}
\end{equation}

The roots of Equation \ref{cf2} are $x_{\pm} = F \pm \sqrt{F^2 - 1}$ which gives a general solution,

\begin{equation}
    \alpha_n = A_{+} x_{+}^n + A_{-} x_{-}^n.
\label{cf3}
\end{equation}

Using the initial conditions we can fix,

\begin{equation}
\begin{split}
    &A_+ = \frac{1}{1-\frac{x_{-}}{x_{+}}},\\
    &A_- = \frac{1}{1-\frac{x_{+}}{x_{-}}},
\end{split}
\label{cf4}
\end{equation}

which gives a solution,

\begin{equation}
    \alpha_n = \frac{x_{+}^{n+1}}{x_{+} - x_{-}} + \frac{x_{-}^{n+1}}{x_{-} - x_{+}}.
\label{cf5}
\end{equation}

Substituting $F=\cos( \theta )$ gives $x_{\pm} = e^{\pm i\theta }$ which we substitute into Equation \ref{cf5},

\begin{equation}
    \alpha_n = \frac{e^{i(n+1)\theta}}{e^{i\theta} - e^{-i\theta}} + \frac{e^{-i(n+1)\theta}}{e^{-i\theta} - e^{i\theta}}.
\label{cf6}
\end{equation}

This simplifies to,

\begin{equation}
    \alpha_n = \frac{\sin((n+1)\theta )}{\sin(\theta )},
\end{equation}

which is the result used in the main paper.

\end{appendices}

\end{document}